\documentclass[aps, prb, twocolumn, superscriptaddress]{revtex4-2}
\usepackage{bm, amsmath, amsfonts, amssymb, ascmac, mathtools, braket}
\usepackage{multirow}
\usepackage{graphicx}
\usepackage{float, color, xcolor}
\usepackage{comment}

\usepackage{bbm}
\usepackage{tabularx}

\usepackage[
pagebackref=false,
colorlinks=true,
linkcolor=blue,
urlcolor=blue,
filecolor=black,
citecolor=red,
pdfstartview=FitV,
pdftitle={},
pdfauthor={},
pdfsubject={},
pdfkeywords={},
pdfpagemode=None,
bookmarksopen=true
]{hyperref}

\newcommand{\ii}{\text{i}}

\begin{document}

\title{Multifractality of many-body non-Hermitian skin effect}

\author{Shu Hamanaka}
\email{hamanaka.shu.45p@st.kyoto-u.ac.jp}
\affiliation{Department of Physics, Kyoto University, Kyoto 606-8502, Japan}

\author{Kohei Kawabata}
\email{kawabata@issp.u-tokyo.ac.jp}
\affiliation{Institute for Solid State Physics, University of Tokyo, Kashiwa, Chiba 277-8581, Japan}

\date{\today}

\begin{abstract}
The non-Hermitian skin effect, anomalous localization of an extensive number of eigenstates induced by nonreciprocal dissipation, plays a pivotal role in non-Hermitian topology and significantly influences the open quantum dynamics. 
However, its genuinely quantum characterization in many-body systems has yet to be developed. 
Here, we elucidate that the skin effect manifests itself as multifractality in the many-body Hilbert space.
This multifractality does not accompany the single-particle skin effect and hence is intrinsic to the many-body skin effect.
Furthermore, we demonstrate that the many-body skin effect coexists with spectral statistics of random matrices, in contrast to multifractality associated with the many-body localization, which necessitates the absence of ergodicity.
We also illustrate multifractality caused by the Liouvillian skin effect in Markovian open quantum systems.
Our work establishes a defining characterization of the non-Hermitian skin effect and uncovers a fundamental relationship between multifractality and ergodicity in open quantum many-body systems.
\end{abstract}

\maketitle

\section{Introduction}

Multifractality emerges ubiquitously in nature~\cite{Mandelbrot-textbook}.
Prime examples in condensed matter physics include multifractal wave functions induced by disorder~\cite{Anderson-58, Abrahams-79, Lee-review}.
Sufficiently strong disorder leads to localization of coherent waves and influences transport properties. 
The interplay of disorder and other system parameters causes localization (Anderson) transitions~\cite{Evers-review}, at which critical single-particle wave functions exhibit multifractality~\cite{Wegner-80, Castellani-86, Schreiber-91, Mudry-96, Evers-00, *Mirlin-00}.
Many-body interactions change the nature of the Anderson localization and lead to the many-body localization~\cite{Gornyi-05, Basko-06, Oganesyan-07, *Pal-10, Huse-review, Abanin-review}.
Multifractality occurs in the depths of many-body localized phases as a result of the intricate structure of the many-body Hilbert space~\cite{Luitz-15, Kravtsov-15, Serbyn-17, Mace-19}.
It also characterizes monitored quantum dynamics~\cite{Zabalo-22, Iaconis-21, Sierant-22}.

Another universal mechanism of localization is the non-Hermitian skin effect~\cite{Lee-16, YW-18-SSH, Kunst-18}.
This phenomenon denotes the extreme sensitivity of the bulk to the boundary conditions due to nonreciprocal dissipation, accompanied by the anomalous localization of an extensive number of eigenstates~\cite{Lee-16, Xiong-18, MartinezAlvarez-18, YW-18-SSH, Kunst-18, Lee-Thomale-19, Liu-19, Lee-Li-Gong-19, Yokomizo-19, Zhang-20, OKSS-20, Denner-21, Okugawa-20, KSS-20, Zhang-22, Kawabata-23, Wang-22}.
Such anomalous localization, not relying on disorder, lacks counterparts in closed systems and is intrinsic to open systems.
The skin effect plays a pivotal role in topological phases of non-Hermitian systems~\cite{Rudner-09, Sato-11, *Esaki-11, Hu-11, Schomerus-13, Leykam-17, Xu-17, Shen-18, *Kozii-17, Gong-18, KSUS-19, ZL-19, Zirnstein-19, Borgnia-19, KSR-21, BBK-review}
and has been experimentally observed in open classical systems of mechanical metamaterials~\cite{Brandenbourger-19-skin-exp, *Ghatak-19-skin-exp}, electrical circuits~\cite{Helbig-19-skin-exp, *Hofmann-19-skin-exp, Zhang-21}, photonic lattices~\cite{Weidemann-20-skin-exp}, and active particles~\cite{Palacios-21}, as well as open quantum systems of single photons~\cite{Xiao-19-skin-exp}, ultracold atoms~\cite{Liang-22}, and digital quantum processors~\cite{Shen-23}.
Recently, beyond band theory, topology and skin effect in non-Hermitian interacting systems have attracted growing interest~\cite{Guo-20, Yoshida-19, Mu-20, Xi-19, Lee-20, Matsumoto-21, Zhang-20Mott, Liu-20, Xu-20, Shackleton-20, CHLee-20, *Shen-22, Yang-21, Yoshida-21, Hyart-22, Orito-22, Alsallom-21, Zhang-Neupert-22, Kawabata-22, Faugno-Ozawa-22, Chen-23, Qin-24, Kim-23, Yoshida-23, Shimomura-24}.
Dynamical signatures of the skin effect have also been investigated within the framework of the quantum master equation~\cite{Song-19, Haga-21, Liu-20PRR, Mori-20, Yang-22, Hamanaka-23, Ehrhardt-24}.
Despite such considerable interest, no genuinely quantum characterizations of the many-body skin effect have been formulated.
For example, several recent works investigated local particle number distributions in real space for non-Hermitian interacting systems~\cite{Mu-20, Lee-20, Zhang-20Mott, Liu-20, Xu-20, Alsallom-21, Zhang-Neupert-22, Kawabata-22, Faugno-Ozawa-22, Kim-23, Yoshida-23}.
However, this approach cannot capture the intricate structure of the many-body Hilbert space or provide quantitative measures of localization.
Consequently, the distinctive role of the skin effect in open quantum many-body systems has remained elusive.

In this work, we elucidate that the skin effect manifests multifractality in non-Hermitian strongly correlated systems, thereby providing a distinctive hallmark of the many-body skin effect.
We also show that the many-body skin effect can coexist with spectral statistics of random matrices.
This contrasts with multifractality associated with the many-body localization, which is incompatible with ergodicity.
In addition to non-Hermitian Hamiltonians, we demonstrate multifractality of the many-body skin effect within the Lindblad master equation.
Our work reveals a defining characteristic of the many-body skin effect and underscores its fundamental role in open quantum many-body systems.

\section{Multifractal scaling}

We consider a normalized wave function $\ket{\psi}$ in a given computational basis $\ket{n}$'s, 
\begin{equation}
\ket{\psi} = \sum_{n = 1}^{\cal N} \psi_n \ket{n} \quad (\psi_n \in \mathbb{C}), 
\end{equation}
with the Hilbert space dimension $\mathcal{N}$.
From the $q$th moments of this wave function, we introduce the $q$th participation entropy as
\begin{equation}
    S_q \coloneqq \frac{1}{1-q} \ln \left( \sum_{n=1}^{\cal N} \left| \psi_n \right|^{2q} \right).
        \label{eq: Sq}
\end{equation} 
For $q=2$, this reduces to the inverse participation ratio 
\begin{equation}
S_2 = - \ln \left( \sum_{n=1}^{\cal N} \left| \psi_n \right|^{4} \right).
\end{equation}

Importantly, the participation entropy $S_q$ quantifies the localization properties in the given Hilbert space.
For perfectly delocalized states, we have $S_q = \ln \mathcal{N}$.
For states localized in a finite region of the Hilbert space, by contrast, $S_q$ no longer depends on $\mathcal{N}$.
Between these two opposite regimes, states can exhibit the intermediate behavior 
\begin{equation}
S_q = D_q \ln \mathcal{N} \quad \left( 0 < D_q < 1 \right), 
\end{equation}
implying that they are extended with nontrivial occupation in the Hilbert space---multifractality.
We employ participation entropy, instead of entanglement entropy, to directly capture such localization properties.
Here, $D_q \coloneqq S_q/\ln \mathcal{N}$ quantifies the effective dimension of the wave function occupying the Hilbert space.

In noninteracting disordered systems, single-particle eigenstates respectively exhibit $D_q = 0$ and $D_q = 1$ in the localized and delocalized phases, between which multifractality $0 < D_q < 1$ can appear concomitantly with the Anderson transitions~\cite{Evers-review}.
In interacting disordered systems, many-body-localized eigenstates can exhibit $0 < D_q < 1$ for substantial disorder~\cite{Mace-19}.
Below, we study the participation entropy $S_q$ in many-body non-Hermitian Hamiltonians and Lindbladians, and demonstrate that the many-body skin effect is distinguished by multifractal dimensions $0 < D_q < 1$.

\section{Model}

To capture a general feature of the many-body skin effect, we study the following nonintegrable non-Hermitian spin chain:
\begin{align}
    H &= \sum_{i=1}^{L} \left[
    \frac{t}{2} \left(
    \left( 1+\gamma \right) \sigma_{i}^{-} \sigma_{i+1}^{+}
    + \left( 1-\gamma \right)   \sigma_{i}^{+} \sigma_{i+1}^{-}
    \right) \nonumber \right. \\
    &\biggl. \qquad \qquad \qquad + J \sigma_{i}^{z} \sigma_{i+1}^{z}
    + g \sigma_{i}^{x} 
    + h  \sigma_{i}^{z} 
    \biggr]
        \label{eq: Ham}
\end{align}
with the real parameters $t, \gamma, J, g, h \in \mathbb{R}$.
Here, $\sigma_i^{x}$, $\sigma_i^{y}$, and $\sigma_i^{z}$ are Pauli matrices at site $i$, and $\sigma_i^{+} \coloneqq \sigma_i^{x} + \ii \sigma_i^{y}$ ($\sigma_i^{-} \coloneqq \sigma_i^{x} - \ii \sigma_i^{y}$) is the spin raising (lowering) operator.
The dimension of the Hilbert space is $\mathcal{N} = 2^{L}$.
In the absence of non-Hermiticity (i.e., $\gamma = 0$), this model reduces to the XXZ model with both longitudinal and transverse fields, a prototypical nonintegrable many-body system realizing the quantum thermal phase~\cite{Kim-Huse-15}.
The non-Hermitian term $\gamma$ describes the asymmetric hopping of the spin magnetization and can be implemented, for example, by continuous monitoring and postselection of the null measurement outcome~\cite{Plenio-review, Daley-review}.
Additionally, such nonreciprocal XX coupling is relevant to the asymmetric simple exclusion process~\cite{Gwa-Spohn-92L, *Gwa-Spohn-92A, Kim-95}.

For $J = g = h = 0$, this model reduces to a non-Hermitian free fermionic model introduced by Hatano and Nelson~\cite{Hatano-Nelson-96, *Hatano-Nelson-97}.
This model is the simplest model subject to the skin effect~\cite{Gong-18, YW-18-SSH, Yokomizo-19}.
In fact, while all single-particle eigenstates form delocalized Bloch waves under the periodic boundary conditions, they are localized at either edge under the open boundary conditions.
In Appendix~\ref{asec: Hatano-Nelson}, we show that these skin modes accompany the vanishing multifractal dimension $D_q = 0$, indicating the perfect localization in the single-particle Hilbert space.

\begin{figure*}[t]
\centering
\includegraphics[width=\linewidth]{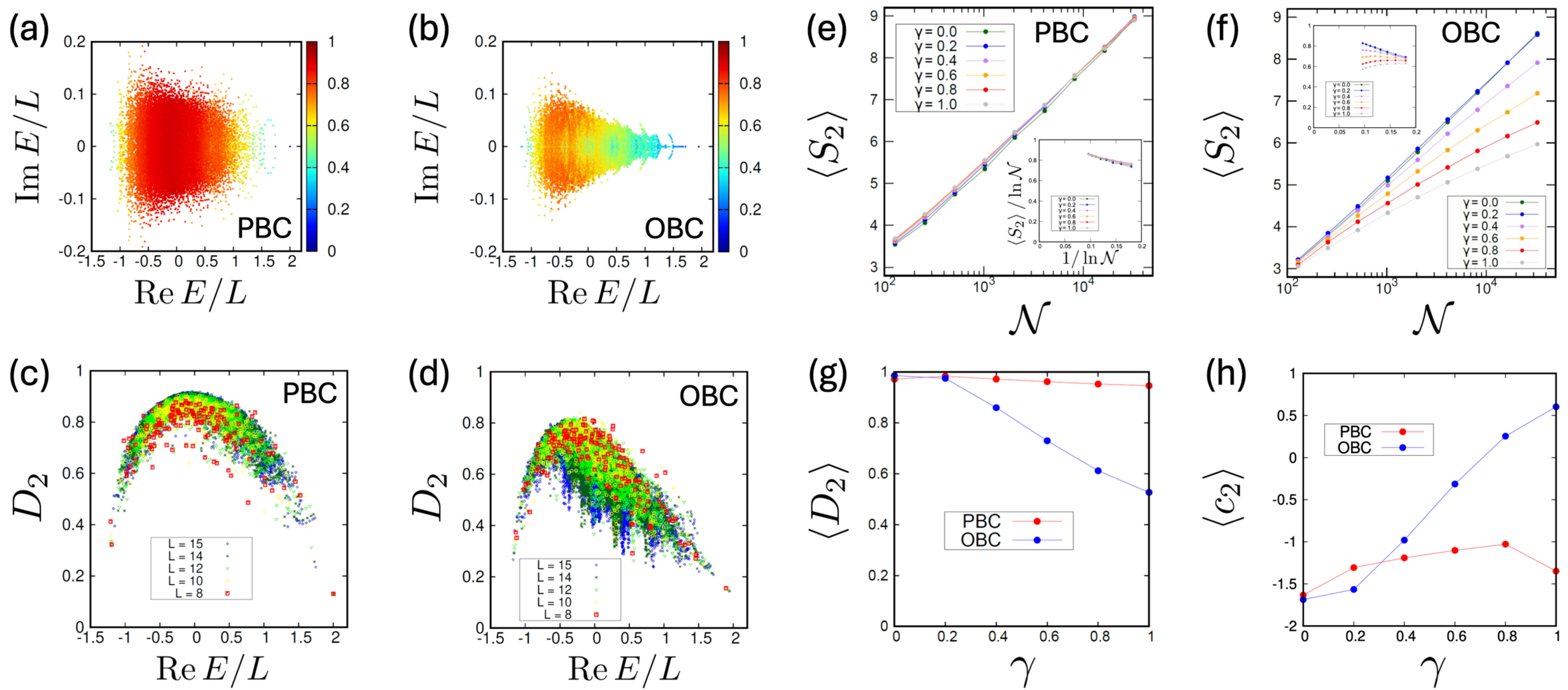}
\caption{Multifractality of the non-Hermitian spin chain in Eq.~(\ref{eq: Ham}) ($t=1/\sqrt{2}$, $J=1$, $g=\left( 5+\sqrt{5}\right)/8$, $h = \left( 1+\sqrt{5}\right)/4$).
(a, b)~Complex spectrum $\left( \mathrm{Re}\,E/L, \mathrm{Im}\,E/L \right)$ scaled by the system length $L=15$ under the (a)~periodic boundary conditions (PBC) and (b)~open boundary conditions (OBC) ($\gamma = 0.8$).
The color bars show the multifractal dimension $D_2 = S_2/\ln \mathcal{N}$ for each right eigenstate.
(c, d)~Multifractal dimensions $D_2$ of individual right eigenstates as a function of $\mathrm{Re}\,E/L$ for the different system lengths $L$ under (c)~PBC and (d)~OBC ($\gamma = 0.8$).
(e, f)~Participation entropy $\braket{S_2}$ averaged over all right eigenstates as a function of the Hilbert space dimension $\mathcal{N}$ under (e)~PBC and (f)~OBC ($L=7,8, \cdots, 15$).
Insets: $\braket{S_2}/\ln \mathcal{N}$ as a function of $1/\ln \mathcal{N}$.
(g, h)~Multifractal dimension $\braket{D_2}$ and its subleading term $\braket{c_2}$ averaged over all right eigenstates as functions of non-Hermiticity $\gamma$ under both PBC (red dots) and OBC (blue dots).}
    \label{fig: Hamiltonian}
\end{figure*}

\section{Multifractality}

Through exact diagonalization, we calculate the complex spectrum and multifractal dimension $D_{q=2}$ for each right eigenstate under both periodic and open boundary conditions [Fig.~\ref{fig: Hamiltonian}\,(a, b)].
While multifractality generally depends on the choice of the computational basis, we here consider the spin configuration.
We choose the model parameters to ensure nonintegrability in the Hermitian limit~\cite{Kim-Huse-15} and investigate generic excited eigenstates instead of special eigenstates including the ground state.
The complex spectrum undergoes substantial changes depending on the different boundary conditions.
While many eigenstates exhibit large multifractal dimensions $D_2 \simeq 1$ under the periodic boundary conditions, their counterparts under the open boundary conditions exhibit much smaller $D_2$, suggesting the skin effect.
Multifractal dimensions $D_2$ quantify the degree of localization dependent on many-body eigenenergies $E$.
Figure~\ref{fig: Hamiltonian}\,(c, d) shows the distributions of multifractal dimensions $D_2$.
Under both boundary conditions, $D_2$ realizes the maximum around the center of the many-body spectrum.
Notably, $D_2$ deviates from unity even under the periodic boundary conditions, which should stem from the locality constraints in a similar manner to Hermitian quantum many-body systems~\cite{Backer-19, *Haque-22}.

As shown in Fig.~\ref{fig: Hamiltonian}\,(a-d), the skin effect influences all many-body eigenstates.
To capture this characteristic feature of the skin effect, we obtain the participation entropy $\braket{S_{q=2}}$ averaged over all right eigenstates as a function of the Hilbert space dimension $\mathcal{N}$ [Fig.~\ref{fig: Hamiltonian}\,(e, f)].
The multifractal dimension averaged exclusively over midspectrum eigenstates shows no significant deviations.
Fitting $\braket{S_2}$ by 
\begin{equation}
\braket{S_2} = \braket{D_2} \ln \mathcal{N} + \braket{c_2}, 
    \label{eq: fitting}
\end{equation}
we obtain the dependence of the average multifractal dimension $\braket{D_2}$ and its subleading term $\braket{c_2}$ on $\gamma$ [Fig.~\ref{fig: Hamiltonian}\,(g, h)].
Whereas $\braket{D_2}$ remains nearly constant under the periodic boundary conditions, it decreases for larger $\gamma$ under the open boundary conditions.
This signifies the stronger skin effect in the many-body Hilbert space.
Importantly, the many-body skin effect does not necessitate disorder, as opposed to the many-body localization.
For small non-Hermiticity $\gamma \lesssim 0.2$, $\braket{D_2}$ seems insensitive to the boundary conditions, implying the absence of the skin effect.
In Appendix~\ref{asec: nonintegrable}, we provide additional numerical results.

Several recent works studied local particle number distributions subject to the skin effect~\cite{Mu-20, Lee-20, Zhang-20Mott, Liu-20, Xu-20, Alsallom-21, Zhang-Neupert-22, Kawabata-22, Faugno-Ozawa-22, Kim-23, Yoshida-23}. 
However, the localization of many-body skin modes should not be captured in real space but in many-body Hilbert space.
The significant difference in multifractal dimensions between the different boundary conditions provides a quantitative measure of the skin effect inherent in non-Hermitian many-body systems.
Additionally, in Appendix~\ref{asec: interacting Hatano-Nelson}, we investigate multifractality in the interacting Hatano-Nelson model~\cite{Zhang-Neupert-22, Kawabata-22, Albertini-96, Fukui-98Nucl, Nakamura-06}.
Despite integrability, the many-body skin effect manifests itself as multifractality, akin to the nonintegrable model in Eq.~(\ref{eq: Ham}).
This further shows the generality of multifractality as a characteristic of the many-body skin effect.

\section{Spectral statistics}

Similarly to many-body skin modes, many-body-localized modes exhibit multifractality.
However, we find a crucial distinction in quantum chaotic behavior, especially spectral statistics.
Several recent works developed measures of chaos in open quantum systems~\cite{Grobe-88, *Grobe-89, Xu-19, Hamazaki-19, Denisov-19, Can-19JPhysA, Hamazaki-20, Akemann-19, Sa-20, JiachenLi-21, GarciaGarcia-22PRX, GJ-23, Sa-23, Kawabata-23SYK}.
While the spectral statistics universally follow the random-matrix statistics in the chaotic regime, they instead follow the Poisson statistics in the integrable regime, providing a diagnosis of the dissipative quantum chaos.

\begin{figure}[t]
\centering
\includegraphics[width=1.0\linewidth]{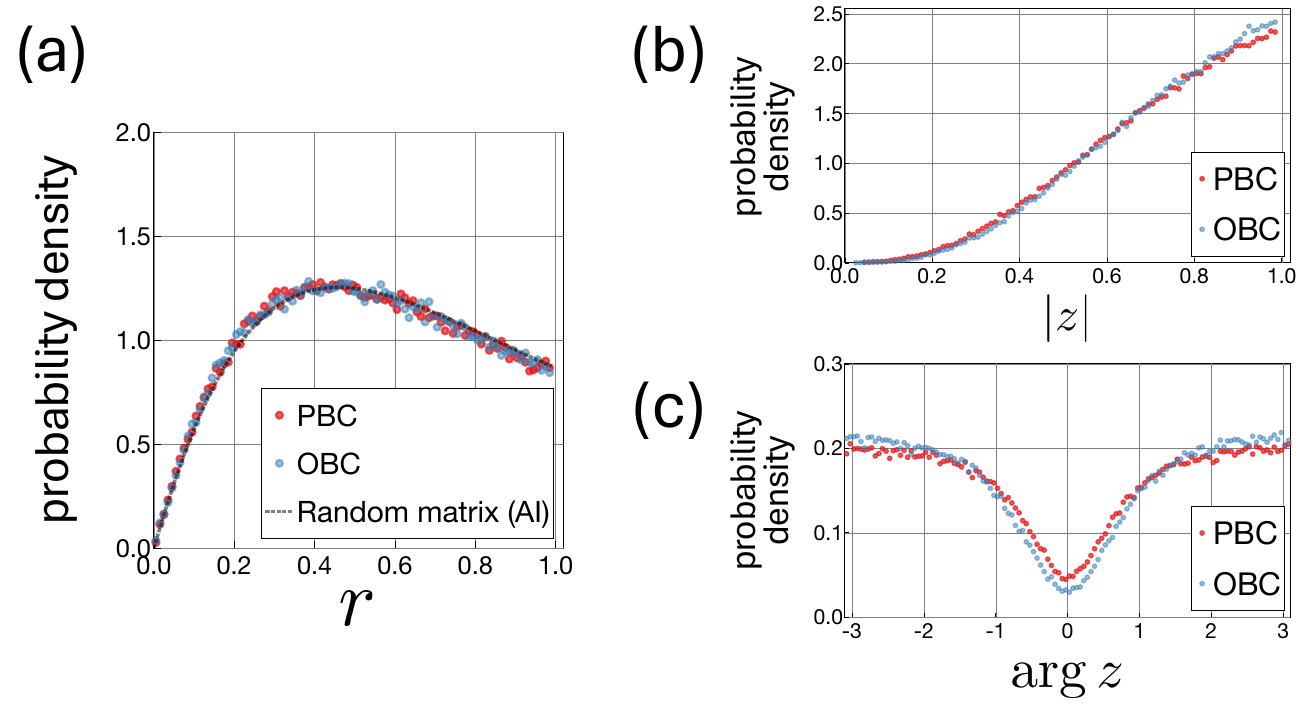} 
\caption{Level-spacing-ratio statistics of the non-Hermitian spin chain in Eq.~(\ref{eq: Ham}) under the periodic boundary conditions (PBC; red dots) and open boundary conditions (OBC; blue dots) ($t = 1/\sqrt{2}$, $\gamma = 0.6$, $J=1$, $g = \left( 5+\sqrt{5} \right)/8$, $h=\left( 1+\sqrt{5} \right)/4$, $L=14$).
All the results are taken away from the spectral edges and the symmetric line, and averaged over $50$ disorder realizations.
(a)~Level-spacing ratio $r$ of singular values.
The averages are $\braket{r} = 0.5297$ for PBC and $\braket{r} = 0.5299$ for OBC.
The black dashed curve is the analytical results for small non-Hermitian random matrices in class AI [i.e., $p \left( r \right) = 27\,( r+r^2 )/4\,( 1+r+r^2 )^{5/2}$; $\braket{r} = 4-2\sqrt{3} \simeq 0.5359$]~\cite{Kawabata-23SVD}.
(b, c)~Level-spacing ratio $z$ of complex eigenvalues for its (b)~absolute value $\left| z \right|$ and (c)~argument $\mathrm{arg}\,z$.
The averages are $\braket{\left| z \right|} = 0.7275$ and $\braket{\left| \cos \mathrm{arg}\,z \right|} = -0.1842$ for PBC and $\braket{\left| z \right|} = 0.7365$ and $\braket{\left| \cos \mathrm{arg}\,z \right|} = -0.2355$ for OBC, while we have $\braket{\left| z \right|} = 0.7381$ and $\braket{\left| \cos \mathrm{arg}\,z \right|} = -0.2405$ for $10^4 \times 10^4$ non-Hermitian random matrices in class A~\cite{Sa-20}.}
	\label{fig: SVD}
\end{figure}

We calculate singular values of the non-Hermitian Hamiltonian in Eq.~(\ref{eq: Ham}) and obtain the distribution of their spacing ratios $r_n$'s [Fig.~\ref{fig: SVD}\,(a)], defined as
\begin{equation}
    r_n \coloneqq \mathrm{min} \left( \frac{s_{n+1} - s_n}{s_n - s_{n-1}}, \frac{s_{n} - s_{n-1}}{s_{n+1} - s_n} \right) \quad \left( 0 \leq r_n \leq 1 \right),
\end{equation}
for an ordered set of singular values $s_n$'s ($n = 1, 2, \cdots, \mathcal{N}$)~\cite{Kawabata-23SVD}.
Here, to break unwanted symmetry, we add a small disordered term $\sum_{i=1}^{L} \varepsilon_i \sigma_i^z \sigma_{i+1}^z$ with a random number $\varepsilon_i$ distributed uniformly in $\left[  -0.1, 0.1 \right]$ for each site $i$, which is expected not to affect multifractality significantly.
Under both periodic and open boundary conditions, the singular-value statistics conform to the statistics of non-Hermitian random matrices, 
indicating the dissipative quantum chaos even in the presence of the skin effect.

We further compute the statistics of complex level-spacing ratios~\cite{Sa-20}, 
\begin{equation}
z_n \coloneqq \frac{E_n^{\rm NN} - E_n}{E_{n}^{\rm NNN} - E_n}
\end{equation}
where $E_n^{\rm NN}$ denotes the nearest-neighbor eigenvalue, and $E_n^{\rm NNN}$ the next-to-nearest-neighbor eigenvalue for each complex eigenvalue $E_n$.
The numerical results conform to the random-matrix statistics, as shown in Fig.~\ref{fig: SVD}\,(b, c).
The slight deviation for the periodic boundary conditions should be due to average translation symmetry.
We note that the non-Hermitian Hamiltonian in Eq.~(\ref{eq: Ham}) respects time-reversal symmetry (i.e., $H^{*} = H$).
Consequently, its spectral statistics follow the random-matrix statistics for class AI~\cite{KSUS-19, Hamazaki-20}.
Time-reversal symmetry also requires the complex spectrum to be symmetric about the real axis [see Fig.~\ref{fig: Hamiltonian}\,(a, b)].

In noninteracting disordered systems at critical points, multifractality accompanies the critical statistics of single-particle spectra that characterize the Anderson transitions~\cite{Evers-review}.
In interacting systems with substantial disorder, multifractality coincides with the Poisson statistics of many-body spectra~\cite{Huse-review, Abanin-review}.
By contrast, we elucidate that the many-body skin effect coexists with the random-matrix statistics even in the presence of multifractality.
This coexistence captures a hallmark of the many-body skin effect and uncovers a distinctive connection between multifractality and ergodicity in open quantum systems.

\begin{figure}[t]
\centering
\includegraphics[width=\linewidth]{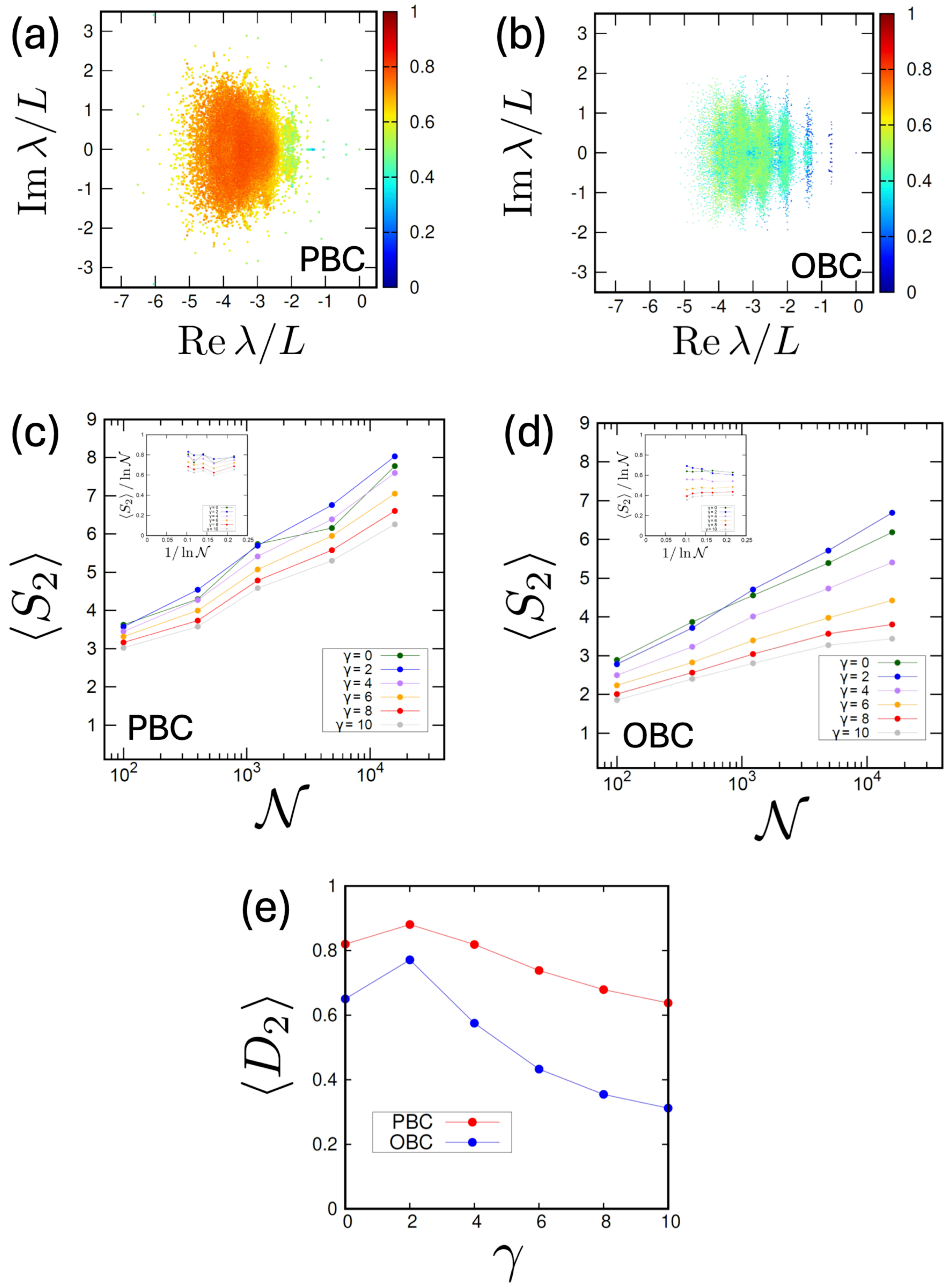} 
\caption{Multifractality of the Liouvillian skin effect ($t=1/\sqrt{2}$, $J=\left( 5-\sqrt{3} \right)/3$, $h = \left( 1+\sqrt{5}\right)/4$).
In both bra and ket spaces, the particle number is chosen as $L/2$ and $\left( L-1 \right)/2$ for even and odd $L$, respectively. 
(a, b)~Liouvillian spectrum $\left( \mathrm{Re}\,\lambda/L, \mathrm{Im}\,\lambda/L \right)$ scaled by the system length $L=9$ under the (a)~periodic boundary conditions (PBC) and (b)~open boundary conditions (OBC) ($\gamma = 6$). 
The color bars show the multifractal dimension $D_2 = S_2/\ln{\mathcal{N}}$ for each right eigenoperator.
(c, d)~Participation entropy $\braket{S_2}$ averaged over all right eigenoperators as a function of the double Hilbert space dimension $\mathcal{N}$ under (c)~PBC and (d)~OBC ($L=5, 6, \cdots, 9$).
Insets: $\braket{S_2}/\ln \mathcal{N}$ as a function of $1/\ln \mathcal{N}$.
(e)~Average multifractal dimension $\braket{D_2}$ as functions of the dissipation strength $\gamma$ under both PBC (red dots) and OBC (blue dots) obtained from the fitting for $L=5, 7, 9$.}
	\label{fig: Lindblad}
\end{figure}

\section{Liouvillian skin effect}
    \label{sec: Liouvillian}

Multifractality accompanies the many-body skin effect even within the quantum master equation.
We investigate Markovian open quantum systems described by $d\rho/dt = \mathcal{L} \left( \rho \right)$ with the Lindbladian~\cite{GKS-76, Lindblad-76, Breuer-textbook}
\begin{equation}
    \mathcal{L} \left( \rho \right) = -\ii\,[H, \rho] + \sum_{n} \left[ L_n \rho L_{n}^{\dag} - \frac{1}{2}\,\{ L_{n}^{\dag} L_{n}, \rho \} \right].
\end{equation}
Here,  $H$ is a Hermitian Hamiltonian for the coherent dynamics, and $L_n$'s are dissipators for the nonunitary coupling with an external environment.
Since $\mathcal{L}$ is a superoperator acting on the density operator $\rho$, its eigenstates are operators.
To study multifractality of these eigenoperators, we double the Hilbert space and map $\mathcal{L}$ and $\rho$ to a non-Hermitian operator and a state, respectively.
Specifically, we transform the bra degree of freedom into an additional ket degree of freedom and map the density operator $\rho = \sum_{ij} \rho_{ij} \ket{i} \bra{j}$ to a pure state $\ket{\rho} = \sum_{ij} \rho_{ij} \ket{i} \ket{j}$ in the double Hilbert space.
Through this operator-state mapping, the Lindblad equation reduces to $d\ket{\rho}/dt = \mathcal{L} \ket{\rho}$ with 
\begin{align}\label{eq:Lindbladian}
    &{\cal L} = -\ii \left( {H} \otimes I - I \otimes {H}^{*} \right) \nonumber \\
    &+ \sum_{n} \left[ {L}_{n} \otimes {L}_{n}^{*} - \frac{1}{2}\,({L}_{n}^{\dag} {L}_{n} \otimes I) - \frac{1}{2}\,(I \otimes {L}_{n}^{T} {L}_{n}^{*}) \right].
\end{align}

We choose the Hermitian Hamiltonian $H$ as Eq.~(\ref{eq: Ham}) with $\gamma = g = 0$.
To realize the skin effect, we consider the nonreciprocal dissipators~\cite{Znidaric-15, Haga-21, Kawabata-23}
\begin{equation}
    L_n = \sqrt{2\gamma} \sigma_n^{-} \sigma_{n+1}^{+} \quad \left( \gamma \geq 0;~n = 1, 2, \cdots, L \right).
\end{equation}
Similar to the non-Hermitian term in Eq.~(\ref{eq: Ham}), these dissipators incoherently push the spin magnetization from the left to the right, inducing the Liouvillian skin effect. 
In the individual bra and ket spaces, this Lindbladian is invariant under $\mathrm{U} \left( 1 \right)$ spin rotation and conserves the spin magnetization $\sum_{i=1}^{L} \sigma_i^z$ [i.e., strong $\mathrm{U} \left( 1 \right)$ symmetry~\cite{Buca-12, Albert-14, Sa-23, Kawabata-23SYK}].
We focus on the half-filled subsector with zero magnetization $\sum_{i=1}^{L} \sigma_i^z = 0$ in both bra and ket spaces.

We exactly diagonalize the Lindbladian $\mathcal{L}$ and obtain the complex spectrum and multifractal dimension $D_{q=2}$ for each right eigenoperator, as shown in Fig.~\ref{fig: Lindblad}.
Depending on whether we impose the periodic or open boundary conditions, the complex spectrum differs substantially, signifying the Liouvillian skin effect. 
For both boundary conditions, the average multifractal dimensions $\braket{D_2}$ deviate from unity, likely due to the limited system size.
Nevertheless, $\braket{D_2}$ under the open boundary conditions is significantly smaller than $\braket{D_2}$ under the periodic boundary conditions, demonstrating multifractality accompanied by the Liouvillian skin effect.
Unlike the previous case for the non-Hermitian Hamiltonian in Eq.~(\ref{eq: Ham}), $\braket{D_2}$ exhibits nonmonotonic dependence on the dissipation strength $\gamma$.
This is because the Lindbladian dissipation incorporates both many-body interactions and nonreciprocity, competing with each other.

\begin{figure}[t]
\centering
\includegraphics[width=\linewidth]{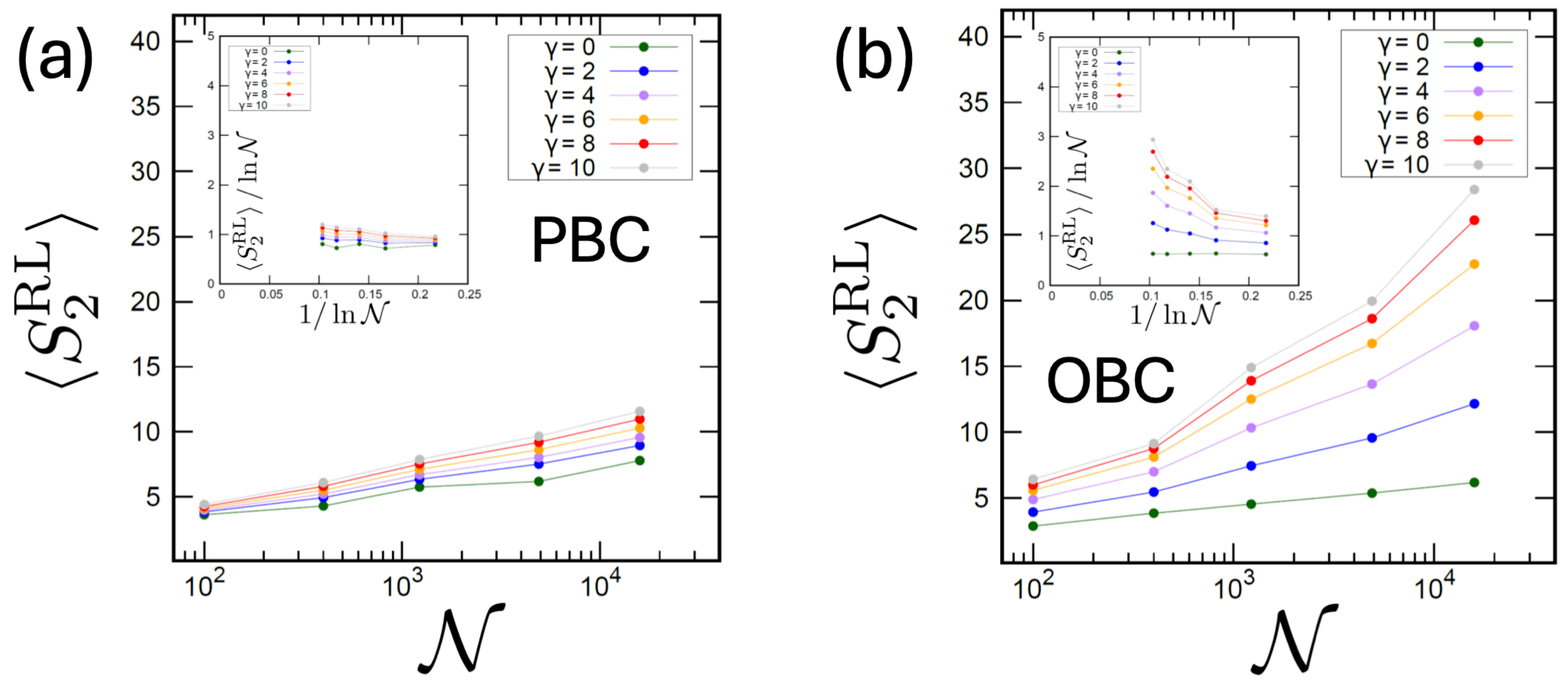} 
\caption{Multifractality of the Liouvillian skin effect ($t=1/\sqrt{2}$, $J=\left( 5-\sqrt{3} \right)/3$, $h = \left( 1+\sqrt{5}\right)/4$).
In both bra and ket spaces, the particle number is chosen as $L/2$ and $\left( L-1 \right)/2$ for even and odd $L$, respectively. 
Participation entropy $\braket{S_2^{\rm RL}}$ averaged over all right and left eigenoperators as a function of the double Hilbert space dimension $\mathcal{N}$ under the (a)~periodic boundary conditions (PBC) and (b)~open boundary conditions (OBC) ($L=5, 6, \cdots, 9$).}
	\label{afig: Lindblad-RL}
\end{figure}

Moreover, we introduce participation entropy and multifractal dimensions based on the combination of right and left eigenoperators, instead of those based solely on right eigenoperators.
We consider right and left eigenoperators of non-Hermitian superoperators,
\begin{equation}
    \ket{r_\alpha} = \sum_{n=1}^{\cal N} r_{n} \ket{n}, \quad \ket{l_\alpha} = \sum_{n=1}^{\cal N} l_{n} \ket{n} \quad \left( r_n, l_n \in \mathbb{C} \right),
\end{equation}
where $\ket{n}$'s form a computational basis, and $\mathcal{N}$ is the dimension of the double Hilbert space.
We normalize these eigenoperators by
\begin{equation}
    \braket{r_\alpha | r_\alpha} = \braket{l_\alpha | l_\alpha} = 1.
\end{equation}
Using both right and left eigenoperators, we introduce the $q$th participation entropy as
\begin{equation}
    S_q^{\rm RL} \coloneqq \frac{1}{1-q} \ln \left( \sum_{n=1}^{\cal N} \left| r_n l_n \right|^{q} \right),
\end{equation} 
and the multifractal dimension as
\begin{equation}
    D_{q}^{\rm RL} \coloneqq \frac{S_q^{\rm RL}}{\ln \mathcal{N}} = \frac{1}{1-q} \frac{1}{\ln \mathcal{N}} \ln \left( \sum_{n=1}^{\cal N} \left| r_n l_n \right|^{q} \right).
\end{equation}

We calculate the participation entropy $\braket{S_{q=2}^{\rm RL}}$ averaged over all right and left eigenoperators (Fig.~\ref{afig: Lindblad-RL}). 
As we increase the double Hilbert space dimension $\mathcal{N}$, the average participation entropy $\braket{S_{2}^{\rm RL}}$ under the periodic boundary conditions grows more slowly than $\braket{S_{2}^{\rm RL}}$ under the open boundary conditions. 
This behavior is qualitatively similar to the behavior of the average participation entropy $\braket{S_{2}}$ defined solely by right eigenoperators (see Fig.~\ref{fig: Lindblad}).
Additionally, as we increase the dissipation strength $\gamma$, the average participation entropy $\braket{S_{2}^{\rm RL}}$ remains almost the same for the periodic boundary conditions but increases for the open boundary conditions.
It is worthwhile to further study the multifractal dimension $D_{q}^{\rm RL}$ defined by both right and left eigenoperators in a more systematic manner.

\section{Discussions}

Despite the considerable recent interest in the non-Hermitian skin effect, its many-body characterization has remained unestablished.
In this work, we have uncovered a connection between two crucial but previously unrelated physical concepts---multifractality and skin effect.
Multifractality provides a defining feature of the skin effect in the many-body Hilbert space.
In contrast to the many-body localization, multifractality due to the many-body skin effect can coexist with the random-matrix spectral statistics.
This lets us revisit a fundamental relationship between multifractality and ergodicity, and reveals a unique role of the skin effect in open quantum systems.

While we have mainly focused on multifractality defined solely by right eigenstates in this work, we also introduce multifractality by the combination of right and left eigenstates in Sec.~\ref{sec: Liouvillian}, 
which may be relevant to the relaxation dynamics~\cite{Haga-21, Mori-20}.
It is worthwhile to further study such different measures of multifractality systematically.
Moreover, recent years have seen various types of nonequilibrium phase transitions in open quantum systems.
In our non-Hermitian spin model in Eq.~(\ref{eq: Ham}), the average multifractal dimension $\braket{D_2}$ under the open boundary conditions remains at unity for $\gamma \lesssim 0.2$ but deviates from unity for $\gamma \gtrsim 0.2$, implying a possible phase transition induced by the interplay between non-Hermiticity and many-body interactions.
It merits further study to determine whether this is a sharp phase transition or a crossover in the infinite-size limit.

\smallskip
{\it Note added}.---After the completion of this work, we became aware of a recent related work~\cite{Gliozzi-24}.

\begingroup
\renewcommand{\addcontentsline}[3]{}
\begin{acknowledgments}
We thank Ryusuke Hamazaki, Takashi Mori, Shoki Sugimoto, and Tsuneya Yoshida for helpful discussions. 
We appreciate the long-term workshop ``Quantum Information, Quantum Matter and Quantum Gravity" (YITP-T-23-01) held at Yukawa Institute for Theoretical Physics (YITP), Kyoto University.
S.H. acknowledges travel support from MEXT KAKENHI Grant-in-Aid for Transformative Research Areas A through the ``Extreme Universe" collaboration (Grant Nos.~21H05182 and 22H05247).
S.H. is supported by JSPS Research Fellow No.~24KJ1445, JSPS Overseas Challenge Program for Young Researchers, and MEXT WISE Program.
K.K. is supported by MEXT KAKENHI Grant-in-Aid for Transformative Research Areas A ``Extreme Universe" No.~24H00945.
\end{acknowledgments}
\endgroup

\appendix

\section{Noninteracting Hatano-Nelson model}
    \label{asec: Hatano-Nelson}

We study the Hatano-Nelson model~\cite{Hatano-Nelson-96, *Hatano-Nelson-97}
\begin{equation}
    \label{eq:single-HN}
    H = 2t \sum_{i=1}^{L} \left[ \left( 1 + \gamma \right) c_{i+1}^{\dag} c_{i} + \left( 1 - \gamma \right) c_{i}^{\dag} c_{i+1} \right],
\end{equation}
where $c_{i}$ ($c_{i}^{\dag}$) annihilates (creates) a fermion at site $i$, $t \in \mathbb{R}$ denotes the average hopping amplitude, and $\gamma \in \mathbb{R}$ denotes the asymmetry of the hopping amplitudes.
The equivalent spin model reads
\begin{equation}
    H = \frac{t}{2} \sum_{i=1}^{L} \left[ \left( 1 + \gamma \right) \sigma_{i}^{-} \sigma_{i+1}^{+} + \left( 1 -\gamma \right) \sigma_{i}^{+} \sigma_{i+1}^{-} \right].
\end{equation}
Here, we calculate multifractal dimensions of right eigenstates in the single-particle Hilbert space for both periodic and open boundary conditions.
The dimension of the single-particle Hilbert space is $\mathcal{N} = L$.

Under the periodic boundary conditions, generic right eigenstates are given as
\begin{equation}
    \ket{\psi_n} = \frac{1}{\sqrt{L}} \sum_{i=1}^{L} e^{\ii k_n i} c_{i}^{\dag} \ket{\rm vac},
\end{equation}
with $k_n \coloneqq 2\pi n/L$ ($n=0, 1, 2, \cdots, L-1$), and the vacuum $\ket{\rm vac}$ of fermions (i.e., $c_i \ket{\rm vac} = 0$ for all $i$).
The $q$th participation entropy in the single-particle Hilbert space is obtained as
\begin{equation}
    S_q = \frac{1}{1-q} \ln \left( \sum_{i=1}^{L} \left| \frac{e^{\ii k_{n} i}}{\sqrt{L}} \right|^{2q} \right) = \ln L
\end{equation}
for arbitrary $q$.
Thus, the eigenstates are perfectly delocalized through the single-particle Hilbert space (i.e., $D_q = 1$).

Under the open boundary conditions, on the other hand, right eigenstates are given as (see, for example, Sec.~SI of the Supplemental Material in Ref.~\cite{Yokomizo-19})
\begin{equation}
    \ket{\psi_n} \propto \sum_{i=1}^{L} \left( \beta^i \sin \left( k_n i \right) \right) c_{i}^{\dag} \ket{\rm vac}, \quad \beta \coloneqq \sqrt{\frac{1+\gamma}{1-\gamma}},
\end{equation}
with $k_n = \pi n/\left( L+1 \right)$ ($n= 1, 2, \cdots, L$).
All of these eigenstates are localized at the right (left) edge for $\gamma > 0$ ($\gamma < 0$), which is a signature of the non-Hermitian skin effect in the single-particle Hilbert space.
For simplicity, we approximate these single-particle eigenstates as
\begin{equation}
    \ket{\psi_n} \simeq \frac{1}{\sqrt{N}} \sum_{i=1}^{L} \left(  \beta e^{\ii k_n} \right)^{i} c_{i}^{\dag} \ket{\rm vac}
\end{equation}
with the normalization constant
\begin{equation}
    N = \sum_{i=1}^{L} \beta^{2i} = \frac{\beta^2 \left( \beta^{2L} - 1 \right)}{\beta^2 - 1}.
\end{equation}
This simplification is essentially the same as the procedure of the non-Bloch band theory~\cite{YW-18-SSH, Yokomizo-19} and should capture the nature of the skin effect.
Alternatively, these states become the exact eigenstates when we add onsite potentials to both edges in an appropriate manner (see, for example, Appendix~D of Ref.~\cite{Kawabata-23}).
Then, the $q$th participation entropy is obtained as
\begin{align}
    S_q &\simeq \frac{1}{1-q} \ln \left( \sum_{i=1}^{L} \left| \frac{\left( \beta e^{\ii k_n} \right)^i}{\sqrt{N}} \right|^{2q} \right) \nonumber \\
    &= \frac{1}{1-q} \ln \left( \left( \frac{\beta^2 - 1}{\beta^2 \left( \beta^{2L} - 1 \right)} \right)^q \frac{\beta^{2q} \left( \beta^{2qL} -1 \right)}{\beta^{2q} - 1} \right) \nonumber \\
    &\simeq \frac{1}{1-q} \ln \left( \frac{\left( \beta^2 - 1 \right)^q}{\beta^{2q} - 1} \right).
\end{align}
In the last approximate equality, we assume $L\to \infty$ and $\beta > 1$ (i.e., $\gamma > 0$).
Thus, $S_q$ does not depend on the system length $L$ for arbitrary $q$, indicating the perfect localization of the eigenstates in the single-particle Hilbert space (i.e., $D_q = 0$).

\section{Nonintegrable non-Hermitian spin chain}
    \label{asec: nonintegrable}

We provide additional numerical results for the nonintegrable non-Hermitian spin chain 
in Eq.~(\ref{eq: Ham}).
To ensure the numerical precision, we perform the same calculations with the two different languages (C++ with LAPACK and Mathematica) and confirm the consistency between the two results.
Additionally, we verify whether we can reproduce the original non-Hermitian Hamiltonian from the complex eigenvalues, along with the right and left eigenvectors, obtained from the numerical diagonalization.

In the absence of non-Hermiticity (i.e., $\gamma = 0$), the eigenstates of Eq.~(\ref{eq: Ham}) exhibit degeneracy, which is lifted upon the introduction of the non-Hermitian term ($\gamma \neq 0$).
In such a case, any linear combination of the degenerate eigenstates remains an eigenstate, making the participation entropy explicitly dependent on the choice of the linear combination.
Given that the degeneracy of eigenstates is lifted in the presence of non-Hermiticity, we here compute the participation entropy in the presence of infinitesimal non-Hermiticity $\gamma$ and regard this result as that for $\gamma = 0$.
We numerically confirm that the participation entropy remains essentially unchanged for $\gamma \lesssim 10^{-4}$ and the difference between $\gamma \to 0$ and small $\gamma \neq 0$ (e.g., $\gamma = 0.2$) is negligible.
It should also be noted that this procedure naturally selects eigenstates.
For example, applying this method to the noninteracting Hatano-Nelson model without non-Hermiticity (i.e., $\gamma = 0$) results in the selection of Bloch states, as the degeneracy for $\gamma = 0$ is lifted.

\begin{figure}[tb]
\centering
\includegraphics[width=\linewidth]{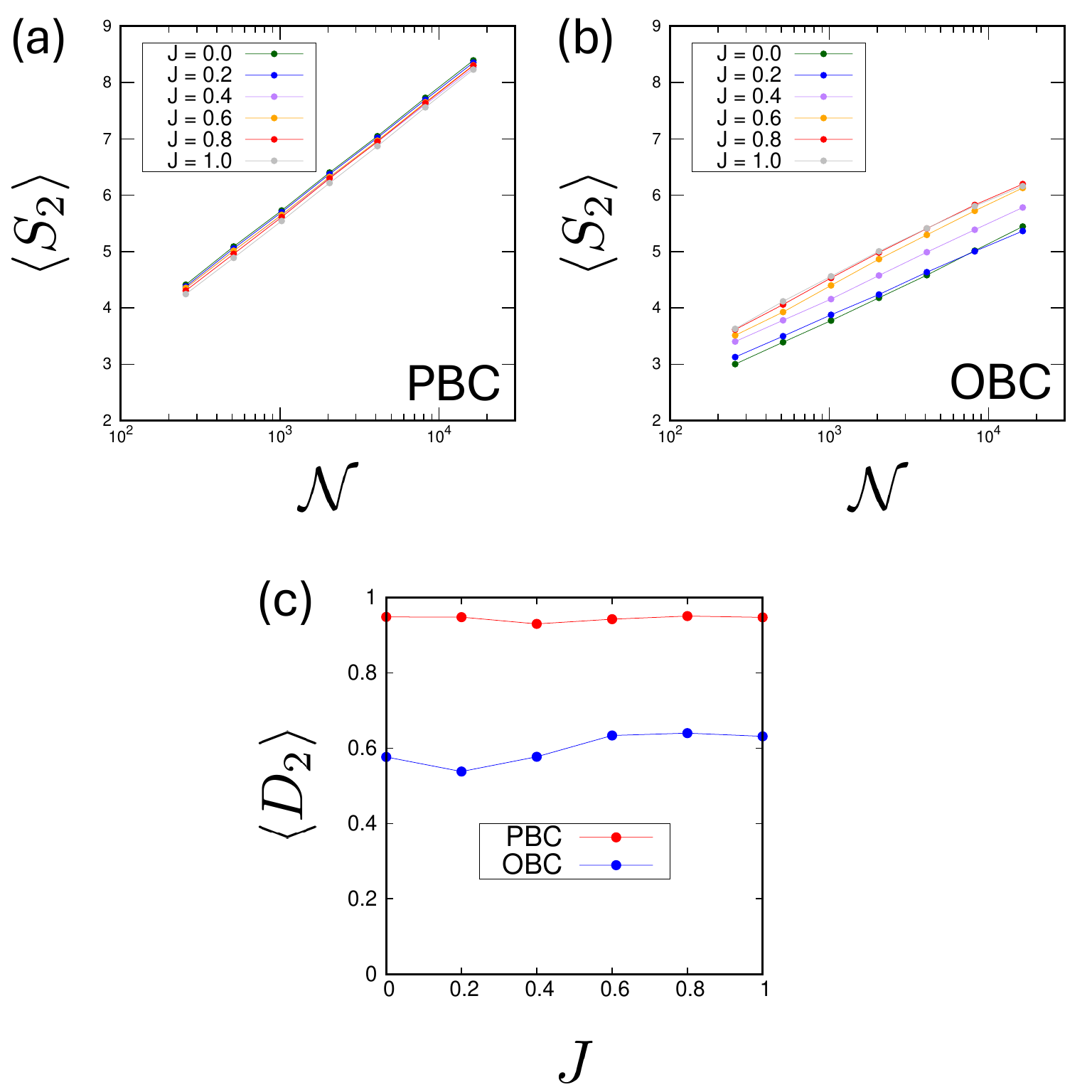} 
\caption{Multifractality of the non-Hermitian spin chain in Eq.~(\ref{eq: Ham}) ($t=1/\sqrt{2}$, $g=\left( 5+\sqrt{5} \right)/8$, $h=\left( 1+\sqrt{5} \right)/4$, $\gamma=0.8$).
(a, b)~Participation entropy $\braket{S_2}$ averaged over all right eigenstates as a function of the Hilbert space dimension $\mathcal{N}$ under the (a)~periodic boundary conditions (PBC) and (b)~open boundary conditions (OBC) ($L= 8, 9, \cdots, 14$).
(c)~Average multifractal dimension $\braket{D_2}$ as functions of the interaction strength $J$ under both PBC (red dots) and OBC (blue dots).}
    \label{afig: S2-Jdep}
\end{figure}

\subsection{Dependence on interaction strength}

In Fig.~\ref{afig: S2-Jdep}, we provide additional numerical calculations of the multifractal dimensions for fixed non-Hermiticity $\gamma = 0.8$ while changing the interaction strength $J$.
Notably, the average multifractal dimensions $\braket{D_2}$ under the open boundary conditions remain nearly unchanged with some nonmonotonic oscillations even if we change $J$.
This should imply that the non-Hermiticity $\gamma$ and many-body interaction $J$ compete with each other, leading to the formation of many-body skin modes.

\begin{figure}[tb]
\centering
\includegraphics[width=\linewidth]{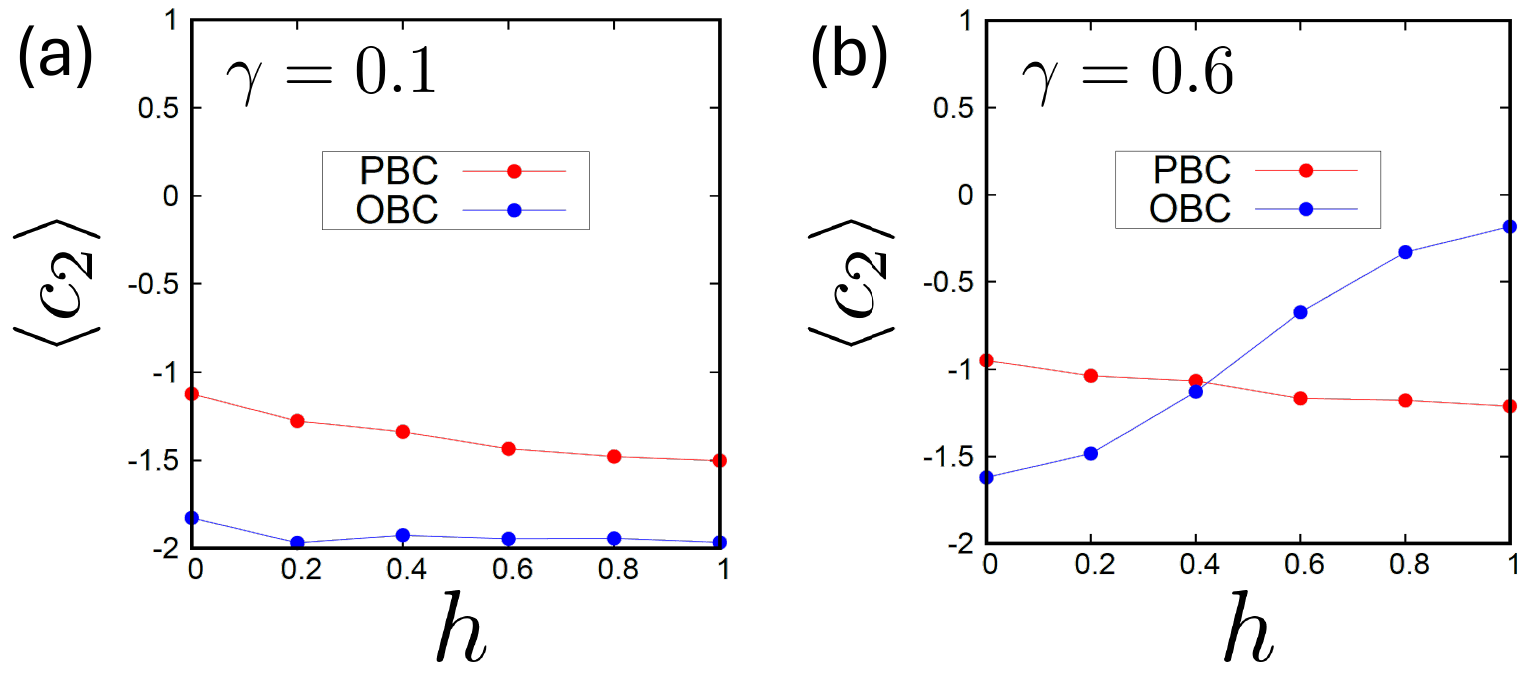} 
\caption{Subleading term $\braket{c_{q=2}}$ of the average participation entropy $\braket{S_{q=2}}$ of the non-Hermitian spin chain in Eq.~(\ref{eq: Ham}) ($t=1/\sqrt{2}$, $J=1$, $g=\left( 5+\sqrt{5} \right)/8$) as a function of the longitudinal field $h$. 
The fitting by $\braket{S_{2}} = \braket{D_2} \ln \mathcal{N} + \braket{c_2}$ is performed up to $L=14$.
We impose the periodic boundary conditions (PBC; red dots) or open boundary conditions (OBC; blue dots).
The non-Hermitian term $\gamma$ is chosen to be (a)~$\gamma = 0.1$ or (b)~$\gamma = 0.6$.}
    \label{afig: S2-hdep}
\end{figure}

\subsection{Dependence on longitudinal field}

In Fig.~\ref{afig: S2-hdep}, we present additional numerical calculations for the subleading term $\braket{c_{q=2}}$ of the average participation entropy $\braket{S_{q=2}}$ for fixed non-Hermiticity $\gamma = 0.1$ or $\gamma = 0.6$ while changing the longitudinal field $h$.
We numerically compute the participation entropy $\braket{S_{q=2}}$ averaged for all right many-body eigenstates and fit it by Eq.~(\ref{eq: fitting}). Under the periodic boundary conditions, the dependence of $\braket{c_2}$ on $h$ is similar for both small non-Hermiticity $\gamma = 0.1$ and large non-Hermiticity $\gamma = 0.6$.
In contrast, under the open boundary conditions, $\braket{c_2}$ appears almost insensitive to $h$ for $\gamma = 0.1$, whereas $\braket{c_2}$ exhibits a more pronounced change for $\gamma = 0.6$.
This significant difference may arise from the absence of the many-body skin effect for small non-Hermiticity, which merits further study.

\begin{figure}[tb]
\centering
\includegraphics[width=\linewidth]{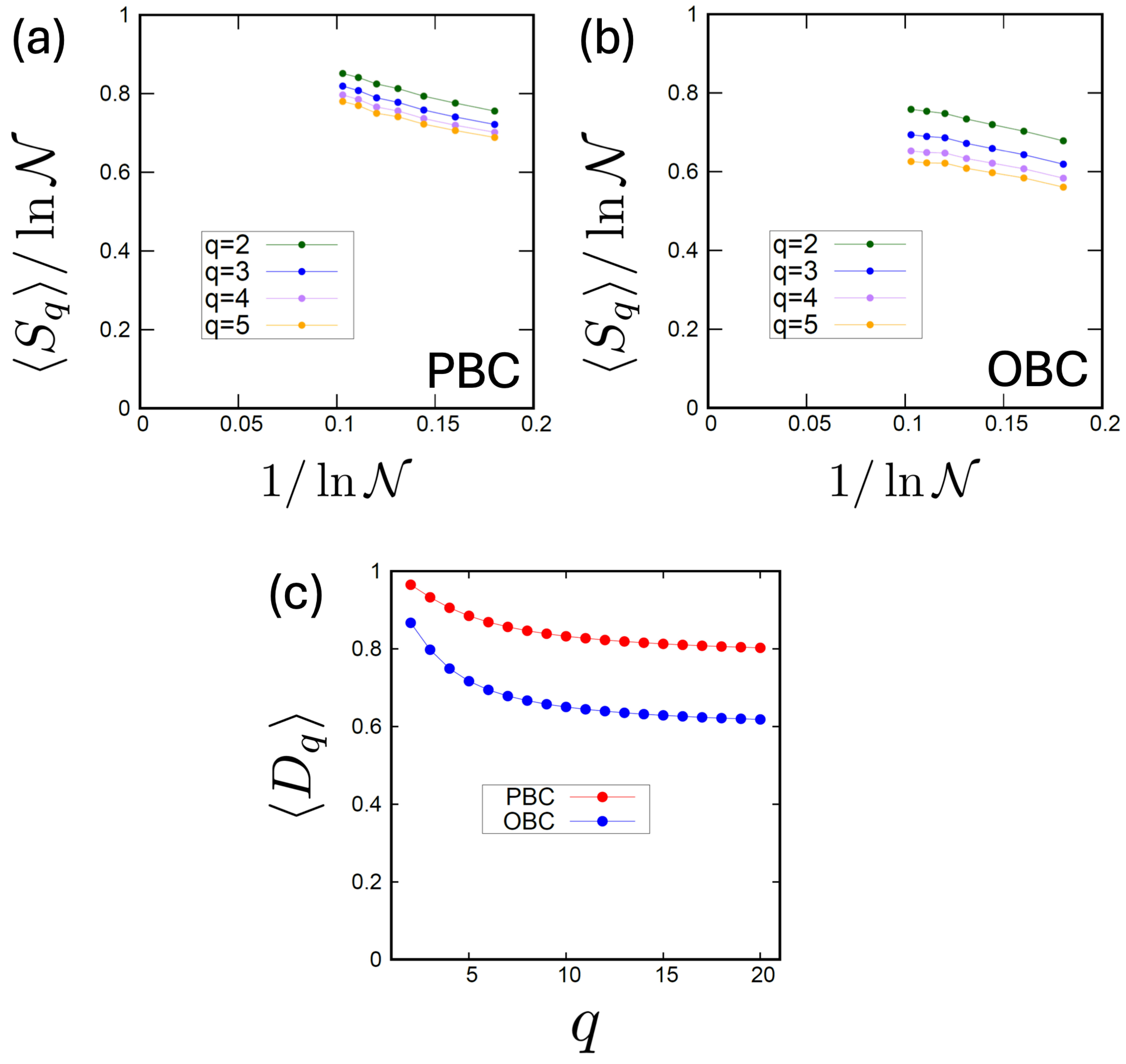} 
\caption{Multifractality of the non-Hermitian spin chain in Eq.~(\ref{eq: Ham}) ($t=1/\sqrt{2}$, $J=1$, $g=\left( 5+\sqrt{5} \right)/8$, $h=\left( 1+\sqrt{5} \right)/4$, $\gamma=0.4$).
(a, b)~$\braket{S_q}/\ln \mathcal{N}$ as a function of $1/\ln \mathcal{N}$ under the (a)~periodic boundary conditions (PBC) and (b)~open boundary conditions (OBC) ($L= 8, 9, \cdots, 14$), where $\braket{S_q}$ denotes the $q$th participation entropy averaged over all right eigenstates, and $\mathcal{N}$ the Hilbert space dimension.
(c)~Average multifractal dimension $\braket{D_q}$ as functions of $q$ under both PBC (red dots) and OBC (blue dots).}
    \label{afig: q-dep}
\end{figure}

\subsection{$q$ dependence}

In Fig.~\ref{afig: q-dep}, we present the dependence of the multifractal dimension $\braket{D_q}$ on $q$.
As shown in Fig.~\ref{afig: q-dep}\,(c), $\braket{D_q}$ exhibits the nontrivial $q$ dependence, which is a defining feature of multifractality.

\begin{figure}[tb]
\centering
\includegraphics[width=\linewidth]{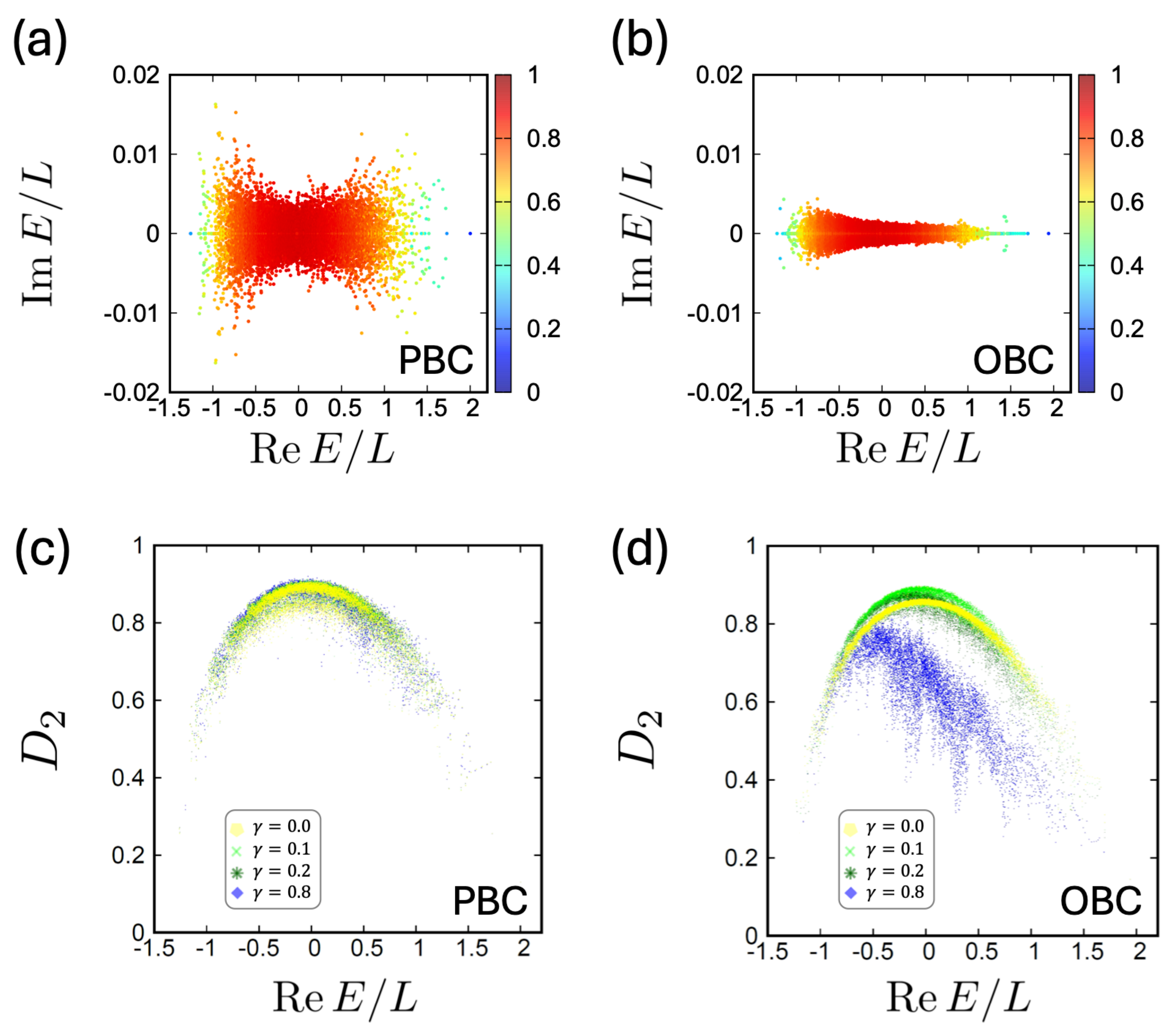} 
\caption{Multifractality of the non-Hermitian spin chain in Eq.~(\ref{eq: Ham}) ($t=1/\sqrt{2}$, $J=1$, $g=\left( 5+\sqrt{5}\right)/8$, $h = \left( 1+\sqrt{5}\right)/4$, $L=14$).
(a, b)~Complex spectrum $\left( \mathrm{Re}\,E/L, \mathrm{Im}\,E/L \right)$ scaled by the system length $L=14$ under the (a)~periodic boundary conditions (PBC) and (b)~open boundary conditions (OBC) ($\gamma = 0.8$).
The color bars show the multifractal dimension $D_2 = S_2/\ln \mathcal{N}$ for each right eigenstate.
(c, d)~Multifractal dimension $D_2$ as a function of $\mathrm{Re}\,E/L$ for different non-Hermiticity $\gamma$ under (c)~PBC and (d)~OBC.}
    \label{afig: small-NH}
\end{figure}

\subsection{Small non-Hermiticity}

In Fig.~\ref{afig: small-NH}, we provide additional numerical calculations of the multifractal dimensions for small non-Hermiticity $\gamma \lesssim 0.2$.
Notably, the multifractal dimensions $D_2$ seem to be largely insensitive to the boundary conditions for $\gamma \lesssim 0.2$, which sharply contrasts with the strong boundary sensitivity for larger non-Hermiticity.
This difference is consistent with the behavior of the average multifractal dimensions $\braket{D_2}$, as shown in Fig.~\ref{fig: Hamiltonian}\,(g).
The boundary insensitivity for small non-Hermiticity can arise from the absence of the non-Hermitian skin effect.
Indeed, when the many-body interaction $J$ is stronger than non-Hermiticity $\gamma$, spins (or more precisely, quasiparticles of spin systems) are bounded by the many-body interaction and cannot exhibit the skin effect.

\begin{figure}[tb]
\centering
\includegraphics[width=\linewidth]{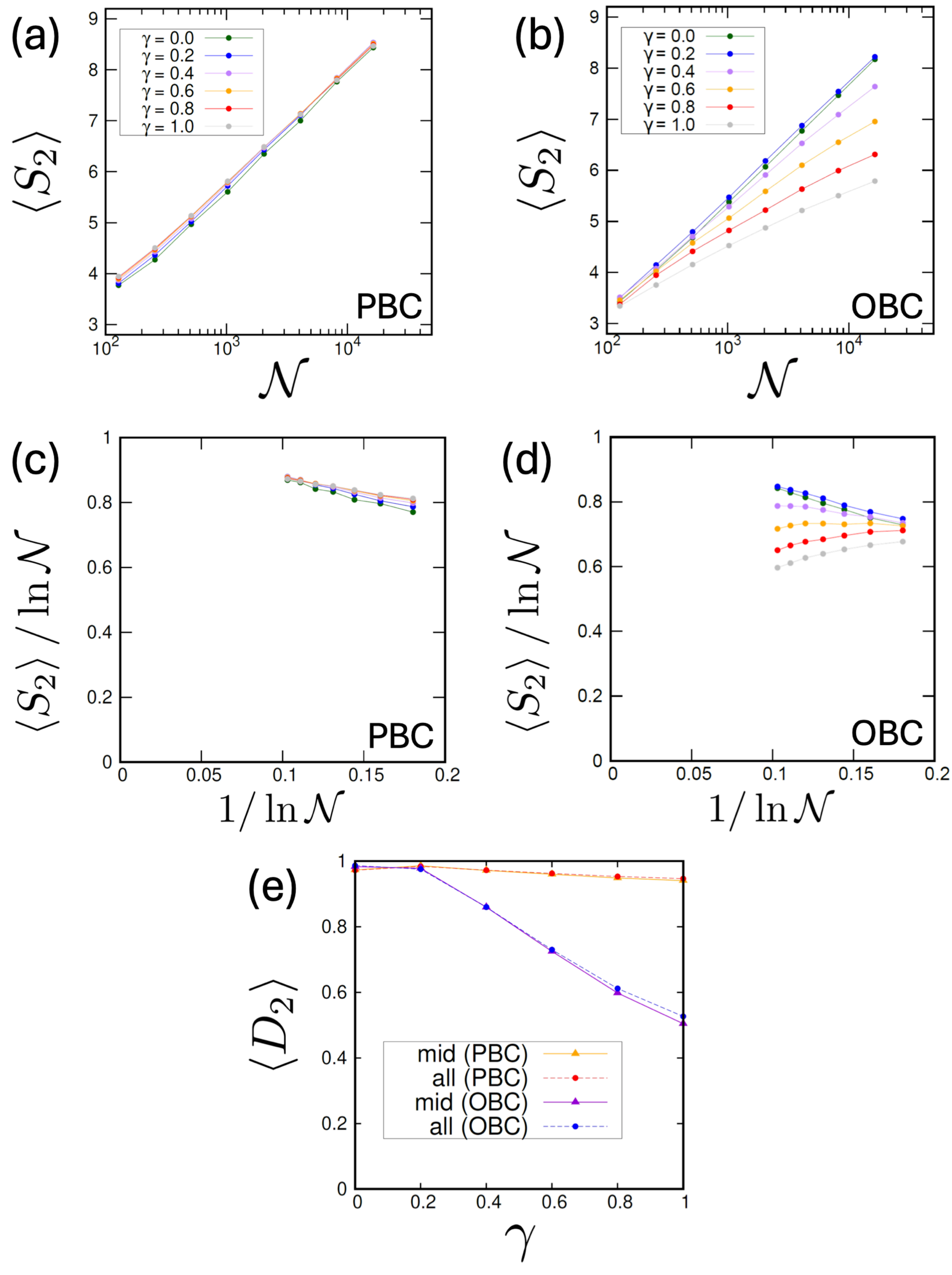} 
\caption{Multifractality of the non-Hermitian spin chain in Eq.~(\ref{eq: Ham}) ($t=1/\sqrt{2}$, $J=1$, $g=\left( 5+\sqrt{5}\right)/8$, $h = \left( 1+\sqrt{5}\right)/4$).
(a, b)~Participation entropy $\braket{S_2}$ averaged over the midspectrum eigenstates with $\left| \mathrm{Re}\,E/L\right| \leq 0.5$ as a function of the Hilbert space dimension $\mathcal{N}$ under the (a)~periodic boundary conditions (PBC) and (b)~open boundary conditions (OBC) ($L=8, 9, \cdots, 14$).
(c, d)~$\braket{S_2}/\ln \mathcal{N}$ as a function of $1/\ln \mathcal{N}$ under the (c)~PBC and (d)~OBC.
(e)~Average multifractal dimension $\braket{D_2}$ for all and midspectrum eigenstates as functions of non-Hermiticity $\gamma$ under the PBC and OBC.}
    \label{afig: midspectrum}
\end{figure}

\subsection{Midspectrum eigenstates}

In Fig.~\ref{afig: midspectrum}, we present additional numerical calculations of the multifractal dimensions averaged exclusively over the midspectrum right eigenstates satisfying $\left| \mathrm{Re}\,E/L \right| \leq 0.5$, instead of considering all the right eigenstates.
Specifically, in Fig.~\ref{afig: midspectrum}\,(e), we compare the multifractal dimensions averaged over all the eigenstates with those over the midspectrum eigenstates.
The results show no significant deviations, thereby validating our use of the average multifractal dimensions to characterize the non-Hermitian skin effect.

\section{Interacting Hatano-Nelson model}
    \label{asec: interacting Hatano-Nelson}

\begin{figure*}[tb]
\centering
\includegraphics[width=1.0\linewidth]{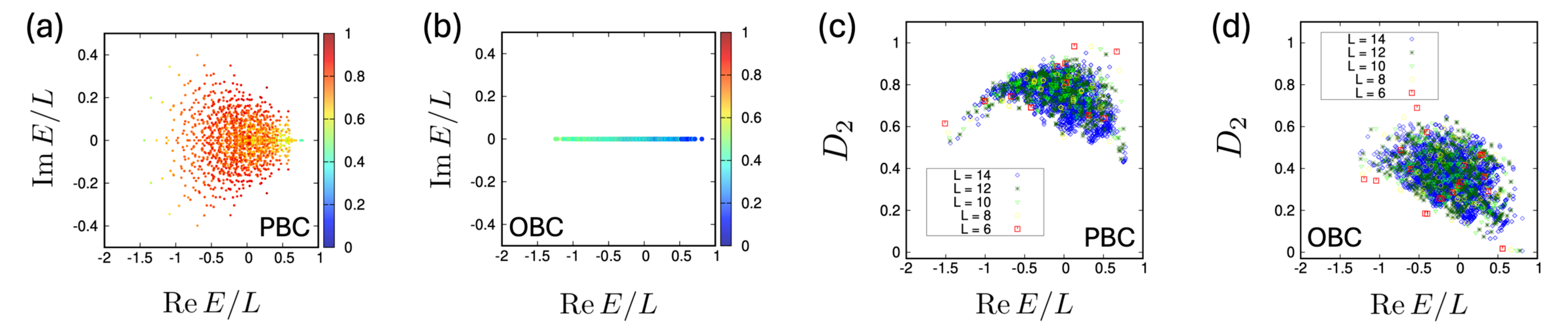} 
\caption{Multifractality of the interacting Hatano-Nelson model ($t=1/\sqrt{2}$, $\gamma = 0.6$, $J=1$).
The particle number is chosen as $L/2$ (i.e., half filling).
(a, b)~Complex spectrum $\left( \mathrm{Re}\,E/L, \mathrm{Im}\,E/L \right)$ scaled by the system length $L=14$ under the (a)~periodic boundary conditions (PBC) and (b)~open boundary conditions (OBC).
The color bars show the multifractal dimension $D_2 = S_2/\ln \mathcal{N}$ for each right eigenstate.
(c, d)~Multifractal dimension $D_2$ as a function of $\mathrm{Re}\,E/L$ for the different system lengths $L$ under (c)~PBC and (d)~OBC.}
	\label{afig: intHN-spectrum}
\end{figure*}

\begin{figure*}[tb]
\centering
\includegraphics[width=0.75\linewidth]{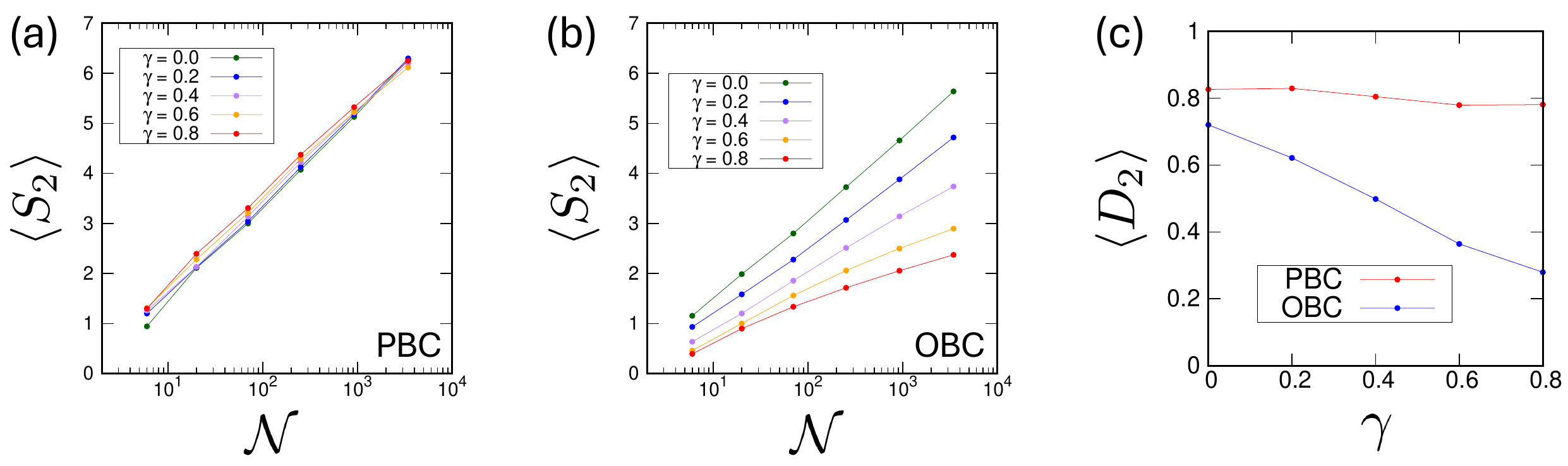} 
\caption{Multifractality of the interacting Hatano-Nelson model ($t=1/\sqrt{2}$, $J=1$).
The particle number is chosen as $L/2$ (i.e., half filling).
(a, b)~Participation entropy $\braket{S_2}$ averaged over all right eigenstates as a function of the Hilbert space dimension $\mathcal{N}$ under the (a)~periodic boundary conditions (PBC) and (b)~open boundary conditions (OBC) ($L=4, 6, \cdots, 14$).
(c)~Average multifractal dimensions $\braket{D_2}$ as functions of non-Hermiticity $\gamma$ under both PBC (red dots) and OBC (blue dots).}
	\label{afig: intHN-multifractal}
\end{figure*}

We consider multifractality of the many-body skin effect in the interacting Hatano-Nelson model~\cite{Zhang-Neupert-22, Kawabata-22}:
\begin{align}
    H &= \sum_{i=1}^{L} \left[ 2t \left( \left( 1 + \gamma \right) c_{i+1}^{\dag} c_{i} + \left( 1 - \gamma \right) c_{i}^{\dag} c_{i+1} \right) \right. \nonumber \\
    &\qquad\qquad\qquad\qquad \left. + 4J c_i^{\dag} c_i c_{i+1}^{\dag} c_{i+1} \right],
\end{align}
where $c_{i}$ ($c_{i}^{\dag}$) annihilates (creates) a fermion at site $i$.
Moreover, $t \in \mathbb{R}$ denotes the average hopping amplitude, $\gamma \in \mathbb{R}$ the asymmetry of the hopping amplitudes, and $J \in \mathbb{R}$ the two-body interaction.
Up to irrelevant constant terms, this non-Hermitian model is equivalent to the XXZ spin chain with the asymmetric XX coupling~\cite{Albertini-96, Fukui-98Nucl, Nakamura-06},
\begin{align}
\label{aeq:integ-HN}
    H &= \sum_{i=1}^{L} \left[
    \frac{t}{2} \left(
    \left( 1+\gamma \right) \sigma_{i}^{-} \sigma_{i+1}^{+}
    + \left( 1-\gamma \right)   \sigma_{i}^{+} \sigma_{i+1}^{-}
    \right) \right. \nonumber \\
    &\qquad\qquad\qquad\qquad\qquad\qquad \left. + J \sigma_{i}^{z} \sigma_{i+1}^{z}
    \right],
\end{align}
which reduces to the non-Hermitian spin chain in Eq.~(\ref{eq: Ham}) with no external fields (i.e., $g=h=0$).
In contrast to Eq.~(\ref{eq: Ham}) with generic parameters, the interacting Hatano-Nelson model is invariant under $\mathrm{U} \left( 1 \right)$ spin rotation around the $z$ axis.
In fact, it respects
\begin{equation}
    \left[ H, \sum_{i=1}^{L} \sigma_i^z \right] = 0.
\end{equation}
Consequently, we study the multifractal scaling in a fixed subspace of the spin magnetization $\sum_{i=1}^{L} \sigma_i^z$.
Below, we assume the even system length $L$ and focus on the half filling (i.e., $\sum_{i=1}^{L} \sigma_i^z = 0$).
It is also notable that this model should be integrable by the Bethe ansatz~\cite{Albertini-96, Fukui-98Nucl, Nakamura-06}. 
This contrasts with Eq.~(\ref{eq: Ham}),
which does not seem to be applicable to the Bethe ansatz and hence should be nonintegrable.
The conformity of the spectral statistics to the random-matrix statistics, shown in Fig.~\ref{fig: SVD}, also corroborates nonintegrability.

We exactly diagonalize the interacting Hatano-Nelson model and obtain the complex spectrum, as well as the multifractal dimension for each right eigenstate [Fig.~\ref{afig: intHN-spectrum}\,(a, b)].
We choose the spin configuration as the computational basis of the participation entropy.
To reduce numerical errors under the open boundary conditions, we first map Eq.~\eqref{aeq:integ-HN} to the Hermitian XXZ chain through the similarity transformation~\cite{Hatano-Nelson-96, *Hatano-Nelson-97}.
We then diagonalize the transformed Hermitian Hamiltonian and return to the original basis, obtaining the participation entropy of the interacting Hatano-Nelson model.
In this procedure, no numerical instability arises since the diagonalization is concerned solely with Hermitian matrices.
Under the periodic boundary conditions, the complex spectrum seems to be more structured than that of Eq.~(\ref{eq: Ham}),
which is also consistent with integrability of the interacting Hatano-Nelson model.
Under the open boundary conditions, the entire many-body spectrum becomes real valued.

Moreover, Fig.~\ref{afig: intHN-spectrum}\,(c, d) shows the distribution of multifractal dimensions as a function of the real part of many-body eigenenergies.
In contrast to Eq.~(\ref{eq: Ham}),
multifractal dimensions no longer exhibit the characteristic behavior in which their peak clearly appears at the center of the many-body spectrum.
Again, this should reflect integrability of the interacting Hatano-Nelson model.

Furthermore, we provide the multifractal scaling in Fig.~\ref{afig: intHN-multifractal}.
In a similar manner to the nonintegrable model in Eq.~(\ref{eq: Ham}),
the average participation entropy $\braket{S_2}$ decreases as non-Hermiticity $\gamma$ increases, showing the stronger many-body skin effect even in the presence of integrability.
We also find that multifractal dimensions deviate from unity even under the periodic boundary conditions.
This should also be a consequence of integrability, which leads to a departure from the random-matrix behavior.

\bibliography{NH_topo.bib}

\end{document}